\documentclass[aps,prl,reprint]{revtex4-1}  

\usepackage{graphicx}
\usepackage{amssymb,amsfonts,amsmath}
\usepackage[rightcaption]{sidecap}
\usepackage{color}

\newcommand{\Sstate}[1]{5\textrm{s}^2\textrm{S}_{1/2}\left( \textrm{m}= #1 1/2 \right)} 
\newcommand{\Dstate}[1]{4\textrm{d}^2\textrm{D}_{5/2}\left( \textrm{m}= #1 5/2 \right)} 
\newcommand{\mKkB}[1]{#1 mK$\cdot k_B$}

\usepackage{times}

\begin{document}

\title{Dynamics of a ground-state cooled ion colliding with ultra-cold atoms}

\author{Z. Meir}
\author{Tomas Sikorsky}
\author{Ruti Ben-shlomi}
\author{Nitzan Akerman}
\author{Yehonatan Dallal}
\author{R. Ozeri}
\affiliation{Department of Physics of Complex Systems, Weizmann Institute of Science, Rehovot 7610001, Israel}

\begin{abstract}
  Ultra-cold atom-ion mixtures are gaining increasing interest due to their potential applications in quantum chemistry, quantum computing and many-body physics.
  Here, we studied the dynamics of a single ground-state cooled ion during few, to many, Langevin (spiraling) collisions with ultra-cold atoms.
  We measured the ion's energy distribution and observed a clear deviation from Maxwell-Boltzmann to a Tsallis characterized by a power-law tail of high energies.
  Unlike previous experiments, the energy scale of atom-ion interactions is not determined by either the atomic cloud temperature or the ion's trap residual excess-micromotion energy. Instead, it is determined by the force the atom exerts on the ion during a collision which is then amplified by the trap dynamics.
  This effect is intrinsic to ion Paul traps and sets the lower bound of atom-ion steady-state interaction energy in these systems.
  Despite the fact that our system is eventually driven out of the ultra-cold regime, we are capable of studying quantum effects by limiting the interaction to the first collision.
\end{abstract}
\maketitle

Since its first inception \cite{MakarovPra, JianLinJmodphys, VladanPrl}, observing quantum mechanical phenomena was the holy grail of hybrid atom-ion research \cite{Zipkesnature, Schmidprl, HudsonFirst, WillitschFirstAtomIon, SmithNew, Rangwala, TakashiPRA}.
The polarization potential between atoms and ions scales as r$^{-4}$ and extends to 100's of nm. This long length-scale interaction can form macroscopic objects \cite{CotePrl} while exhibiting quantum features such as Feshbach \cite{IdziaszekPra,TomzaYbLi} and shape resonances \cite{DulieuNJP} at sufficiently low temperatures.
Ultra-cold atom-ion systems are appealing new platforms for performing quantum computation  \cite{GerritsmaDoubleWell,TommasoQuantumGate} and many-body physics \cite{GerritsmaSolidState}.
However, all experiments so far relied on sympathetic cooling of the ion by the atomic cloud and were limited to above mK energy scale.

The interplay between the ion's radio-frequency (rf) trap oscillating electric fields and sympathetic cooling has been known for a long time \cite{MajorSympatheticCooling}. In particular, it was theoretically shown that sympathetic cooling of the ion results in a non-thermal energy distribution with a power-law tail, the magnitude of which depends on the atom-ion mass ratio and trap parameters \cite{DeVoePowerLaw,Zipkesnewjphys,HudsonSympatheticCooling}.
This phenomenon is closely related, among other examples, to anomalous diffusion in optical lattices \cite{lutz2003} and is well described by non-Maxwellian statistics which was introduced by Tsallis \cite{Tsallis1988}. If the ion is sufficiently lighter than the cooling atoms its mean energy diverges and collisions eventually result in ion loss from trap.

In stable mixtures the mean steady-state energy of the ion is proportional, albeit with potentially a large amplification factor, to the energy of a single collision at the trap center which is determined by, e.g., residual excess-micromotion (EMM) \cite{Zipkesnewjphys} or the atoms' temperature \cite{HudsonSympatheticCooling}. It is therefore interesting to ask what would determine the steady state temperature and ion dynamics when the ion is initialized in the ground state of the trap and both ion's EMM and atoms' temperature are negligible? Recently, the energy involved in a single collision was calculated to be determined by the atom-ion attraction during collision which pulls the ion away from the trap minimum into finite rf regions of the trap \cite{CetinaMM}. Furthermore, the quantum dynamics of ultra-cold atom-ion collisions was calculated and has shown significant heating that depends on the trap parameters used \cite{IdziaszekMM}.

In this experiment, we studied the dynamics of an ion, initialized in the ground-state of all trap modes, during collisions with ultra-cold atoms and negligible EMM, thus investigating the fundamental limits to the temperature of atom-ion mixtures in Paul traps. The species we used are $^{87}$Rb atoms and $^{88}$Sr$^+$ ion which have almost equal masses. This choice of masses amplifies the deviation from Maxwell-Boltzmann to a power-law energy distribution which was not observed in experiments before. Our results show a clear deviation from Maxwell-Boltzmann to Tsallis energy distribution with a pronounced power-law tail. Moreover, the heating mechanism in our system is indeed seeded by the pulling of the ion from the trap center and is in good agreement with molecular dynamics simulations.

During a collision the atom is polarized by the electric field of the ion, leading to an attractive potential $V\left(r\right)=-C_4/2r^4$. Here, $r$ is the atom-ion separation and $C_4$, which is proportional to the polarizability of the atom, characterizes the interaction strength. Classically, atom-ion collisions are divided into glancing and Langevin (spiraling) collisions by the impact parameter $b_c$=$\left(2C_4/E_{col}\right)^{1/4}$, where $E_{col}$ is the collision energy. Glancing collisions, which have a larger impact parameter than $b_c$, are purely elastic and involve only very small momentum transfer. In contrast, Langevin collisions involve large momentum transfer and can also lead to inelastic processes such as spin exchange or relaxation \cite{Spin_Kohl}, charge exchange \cite{ChargeExchange_Kohl}, molecule formation \cite{WillitschFirstAtomIon} and three-body recombination \cite{ThreeBody_denschlag}.
While the loss of atoms from the trap in the presence of an ion is dominated by glancing collisions \cite{Zipkesnewjphys} we expect the heating of the ion to be dominated by the large momentum transfer of Langevin collisions, the rate of which,
$\Gamma_{L}=2\pi n_{at} \sqrt{\textrm{C}_4/\mu}$, is energy independent. For our experimental parameters (reduced mass $\mu=\left(m_{at}m_{ion}\right)/\left(m_{at}+m_{ion}\right)\approx m_{ion}/2$ and atomic peak density $n_{at}\approx1.2\cdot10^{12}$ cm$^{-3}$) the mean time between Langevin collisions is $t_{L}=1/\Gamma_{L}=0.35$ ms.

During collisions the ion's energy distribution develops a power-law tail \cite{DeVoePowerLaw,Zipkesnewjphys,HudsonSympatheticCooling,multipoletrap}. We use the Tsallis distribution which is a generalization of the thermal Maxwell-Boltzmann distribution to fit both our simulation and experiment results \cite{Tsallis1988},
\begin{equation}\label{Eq:Tsallis}
P\left(E\right)=A_n\frac{E^2}{\left(1+\frac{E}{n k_{B} T}\right)^{n}}.
\end{equation}
Here, $A_n=(n-3)(n-2)(n-1)/\left(2(nk_{B} T)^3\right)$ is a normalization factor, $k_B$ is the Boltzmann constant, $E$ is the ion's energy and $T$ and $n$ are parameters of the distribution. In the literature, Tsallis functions are usually defined with the q-parameter, $q_{_T}$. Here we define $n=1/\left(q_{_T}-1\right)$ such that in the limit of $n$$\to$$\infty$ ($q_{_T}$$\to$1) the distribution in Eq. \ref{Eq:Tsallis} becomes a thermal distribution of a 3D harmonic oscillator: $P\left(E\right)\propto E^2 e^{-E/k_B T}$. For smaller n-values the distribution exhibits power-law asymptotic tail: $P\left(E\right)\sim E^{2-n}$. It is also important to notice that the distribution is non-normalizable for $n$$\leq$3 and the distribution mean diverges for $n$$\leq$4.

Our experiment is designed to overlap ultra-cold $^{87}$Rb atoms ($\sim$5 $\mu$K) trapped in a cross dipole trap with ground-state cooled $^{88}$Sr$^+$ ion ($\bar n$$<$0.1 in all three modes of motion) trapped in a linear segmented Paul trap. The ion's EMM is routinely evaluated and compensated ($E_{EMM}$$<$0.5 mK$\cdot k_B$) using side-band spectroscopy on a narrow optical transition. Using optical-pumping we initialize the ion in the $\Sstate{-}$ Zeeman sub-level. The atoms are prepared in the F=1 hyperfine manifold of their ground electronic state and are not polarized \cite{Methods}.
We typically overlap 20,000 atoms with the ion for a variable interaction time ranging from 0.5 ms to several seconds, at the end of which the atoms are released from the dipole trap.
Following interaction we measure the ion's energy. Different interaction times lead to different ion energies and therefore two different ion-thermometry methods were used. Following short interaction times and with energies up to few mK, carrier Rabi spectroscopy \cite{DidiRMP} of the narrow electric quadrupole transition was used. For longer interaction times, we used the Doppler re-cooling (DRC) method \cite{ReCooling} on a strong dipole allowed transition.

Each experimental run, typically lasting few seconds of atom cloud preparation, transport and atom-ion interaction, ends with ion interrogation. Since atom-ion collisions lead to spin de-polarization \cite{Spin_Kohl}\footnote{Work in progress}, we used a short optical pumping (OP) pulse to transfer the population back to the $\Sstate{-}$ state before performing the Rabi spectroscopy. Immediately after, we shine a pulse of light resonant with the $\Sstate{-}\rightarrow\Dstate{-}$ narrow quadrupole transition for a duration $t_R$, after which we determine whether the ion was shelved to the meta-stable D-state using state selective fluorescence on the $\textrm{S}_{1/2}\rightarrow\textrm{P}_{1/2}$ transition. The shelving probability is given by,
\begin{equation}\label{Eq:RabiFlop}
	P_D\left(t_R\right)=\sum_{\mathbf{n}}P\left(\mathbf{n}\right)\sin^2 \left(\Omega_{\mathbf{n}}t_R\right).
\end{equation}
Here, $\mathbf{n}=\left(n_x,n_y,n_z\right)$ is the ion's harmonic oscillator state. $\Omega_{\mathbf{n}}=\Omega_0\prod_i e^{-\frac{\eta_i^2}{2}} L_{n_i}\left(\eta_i^2\right)$ is the carrier Rabi frequency with $\Omega_0$ the bare Rabi frequency, $L_{n_i}\left(x\right)$ is the Laguerre polynomial of degree $n_i$ and $\eta_i$ is the Lamb-Dicke parameter of the i-th mode.
The ion's total energy is $E=\sum_{i=x,y,z}(\hbar \omega_i n_i + 1/2)$ where $\omega_i/2\pi$ is the i-th mode frequency and $\hbar$ is the reduced Planck constant. The ion's energy distribution $P\left(E\right)$ is given by Eq. \ref{Eq:Tsallis}.

\begin{figure*}
	\centering
	\includegraphics[width=\linewidth]{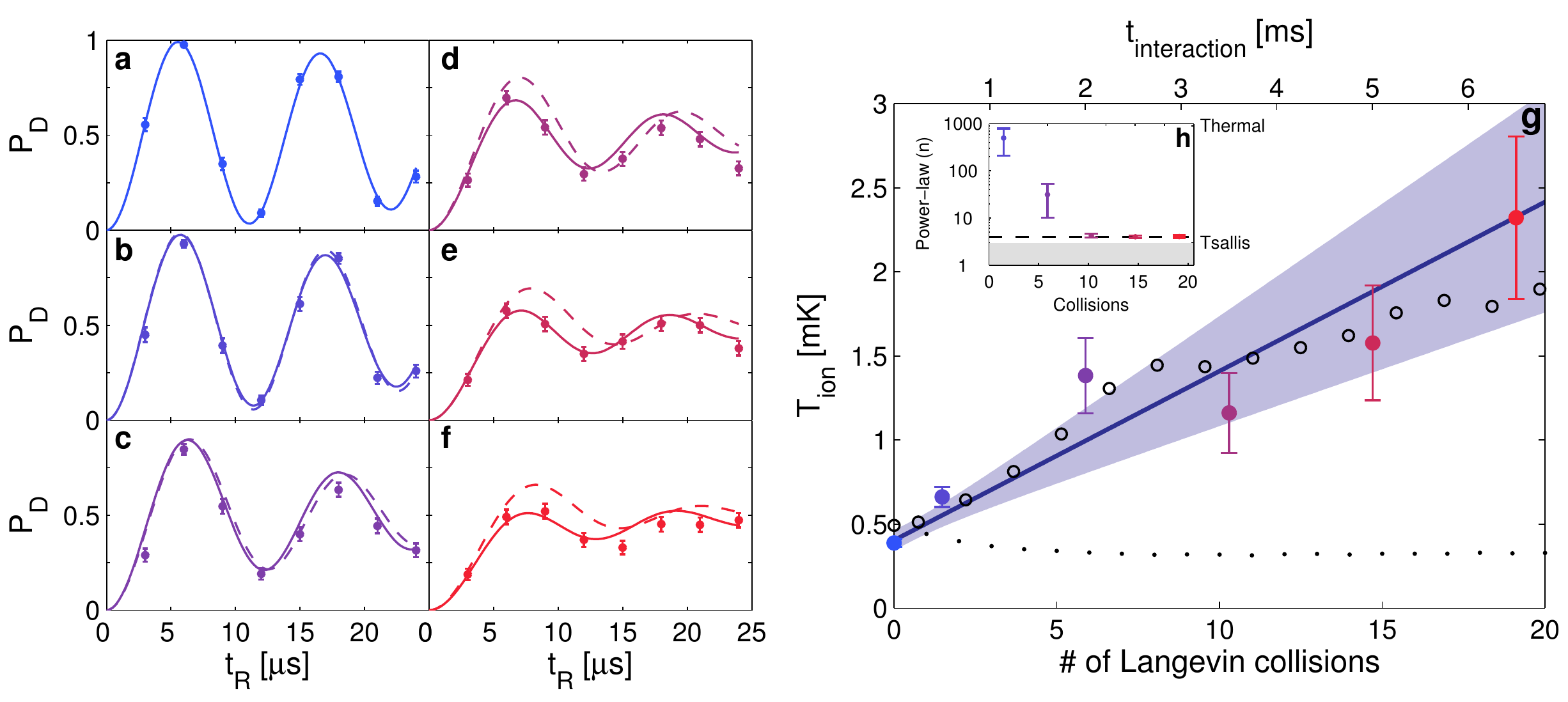}\\
	\caption{\textbf{Carrier Rabi spectroscopy for few atom-ion Langevin collisions.} \textbf{a-f)} Each graph corresponds to a different interaction time (0,0.5,2,3.5,5,6.5 ms for graphs a-f). We scan the shelving pulse time $t_R$, and measure the shelving probability $P_D$. Each data point corresponds to 170 experiments. Error-bars are binomial distribution standard-deviation. We fit the data using Eq. \ref{Eq:RabiFlop} together with the Tsallis distribution (Eq. \ref{Eq:Tsallis}). We extract the distribution free parameters ($T$ and $n$) using maximum likelihood estimation \cite{Methods}. The resulting curve is shown in solid lines. Dashed lines show fit of our data to a thermal distribution.
	\textbf{g)} The ion's temperature ($T_{Ion}=T n / \left(n-2\right)$) increases linearly with a rate of 296(37) $\mu$K$/$ms which is equivalent to 100(13) $\mu$K per collision. Error-bars are one sigma standard deviation. Shaded area represents linear fit confidence bounds (one sigma). Open circles are the results of a simulation which takes into account the reaction of the polarization potential on the ion's position. Black dots are simulation results taking into account only hard-sphere collisions.
	\textbf{h)} Ion's power-law parameter, $n$. The ion's energy distribution starts with $n$$\gg$$1$ consistent with a Maxwell-Boltzman distribution, and converges to $n$=4.0(2) after $\sim$10 collisions. For $n$$>$$10$ Thermal and Tsallis distributions are almost indistinguishable as can be seen in \ref{Fig:figure2}a-c. The grey shaded area represents the non-normalizable region of the distribution ($n$$<$$3$). The dashed line represent the threshold ($n$$=$$4$) for mean energy divergence.}
	\label{Fig:figure2}
\end{figure*}

The experimental results for atom-ion interaction times lasting up to 6.5ms, which correspond to up to 20 Langevin collisions on average, are shown in figures \ref{Fig:figure2}a-f. We scanned the pulse duration, $t_R$, and fitted the measured shelving probability (shown by the filled circles) to Eq. \ref{Eq:RabiFlop} using the distribution of Eq. \ref{Eq:Tsallis}. We estimated the distribution free parameters, $T$ and $n$, using maximum likelihood \cite{Methods} (the best fit is shown by solid lines). As seen, the ion heats up due to collisions with the atoms, and as it does, the contrast of the flopping curve decays due to incoherent sum of contributions from different motional states. As seen in Fig. \ref{Fig:figure2}h, the energy distribution changes from thermal ($n$$\gg$$1$) to a power-law distribution with $n$=4.0(2) over the course of several collisions. A comparison to the best fitted thermal distribution is shown by the dashed lines in figures \ref{Fig:figure2}a-f. As seen, a thermal distribution fails to faithfully explain our observations.

Once the energy distribution of the ion has been determined, we examine the rate of ion heating as a function of the interaction time. Since the ion's mean energy is not well defined for this power-law we characterize the distribution using the most probable energy, $E_{mode}$,
\begin{equation}\label{Eq:Tmode}
k_B T_{Ion}=E_{mode}/2=k_B T n / \left(n-2\right),
\end{equation}
which we will hereafter refer to as the ion's temperature. Note that for thermal distribution $T_{Ion}=T$.
The temperature of the ion is shown by the filled circles in Fig. \ref{Fig:figure2}g. The heating is linear with a rate of 296(37) $\mu$K$/$ms which corresponds to, on average, 100(13) $\mu$K per collision. After 6.5 ms the ion's temperature exceeds 2 mK and the carrier thermometry losses sensitivity. At this point the ion has already heated up significantly beyond its EMM energy. It is important to note that even though our ion is initialized in the ground-state the first point in Fig. \ref{Fig:figure2}g is significantly higher than the ground-state temperature. This is due to beam-pointing instability during the $>$day data acquisition time in this experiment. The beam pointing only affects the cold temperature points at which Rabi flops have high contrast (Fig. \ref{Fig:figure2}a).

To gain better understanding of the heating dynamics, we compare our results to a molecular-dynamics simulation \cite{Methods}. First, we use a simulation similar to Ref. \cite{Zipkesnewjphys} which only takes into account hard-sphere collisions and therefore is only affected by EMM and the temperature of the atoms (black dots in Fig. \ref{Fig:figure2}g). Here, the ion equilibrates with the residual EMM energy (set to 0.5 mK$\cdot k_B$ in the simulation). The resulting ion's energy distribution is also power-law with $n$=3.83. However, the ion's steady-state temperature of $\sim$0.35 mK is almost order of magnitude lower than the last data point of our experimental results which shows no steady-state behavior in the measured regime.
As a second step, we add to our simulation the polarization force between the atom and ion and calculate the particles trajectories in a similar fashion to Ref. \cite{CetinaMM} (empty circles in Fig. \ref{Fig:figure2}g). In this case, the simulation faithfully reproduces our experimental results. The ion's temperature increases linearly at the experimental rate to 2 mK and the distribution power-law, $n$, converges to $n$=3.8.
As seen, in the absence of EMM and negligible atomic temperature, the ion dynamics is dominated by the reaction of the polarization potential on the ion, pulling it away from the trap center into finite RF-regions in the trap. This is the first observation of atom-ion collision dynamics which is not determined by the atom's temperature or the ion's EMM.

\begin{figure}
	\centering
	\includegraphics[width=\linewidth]{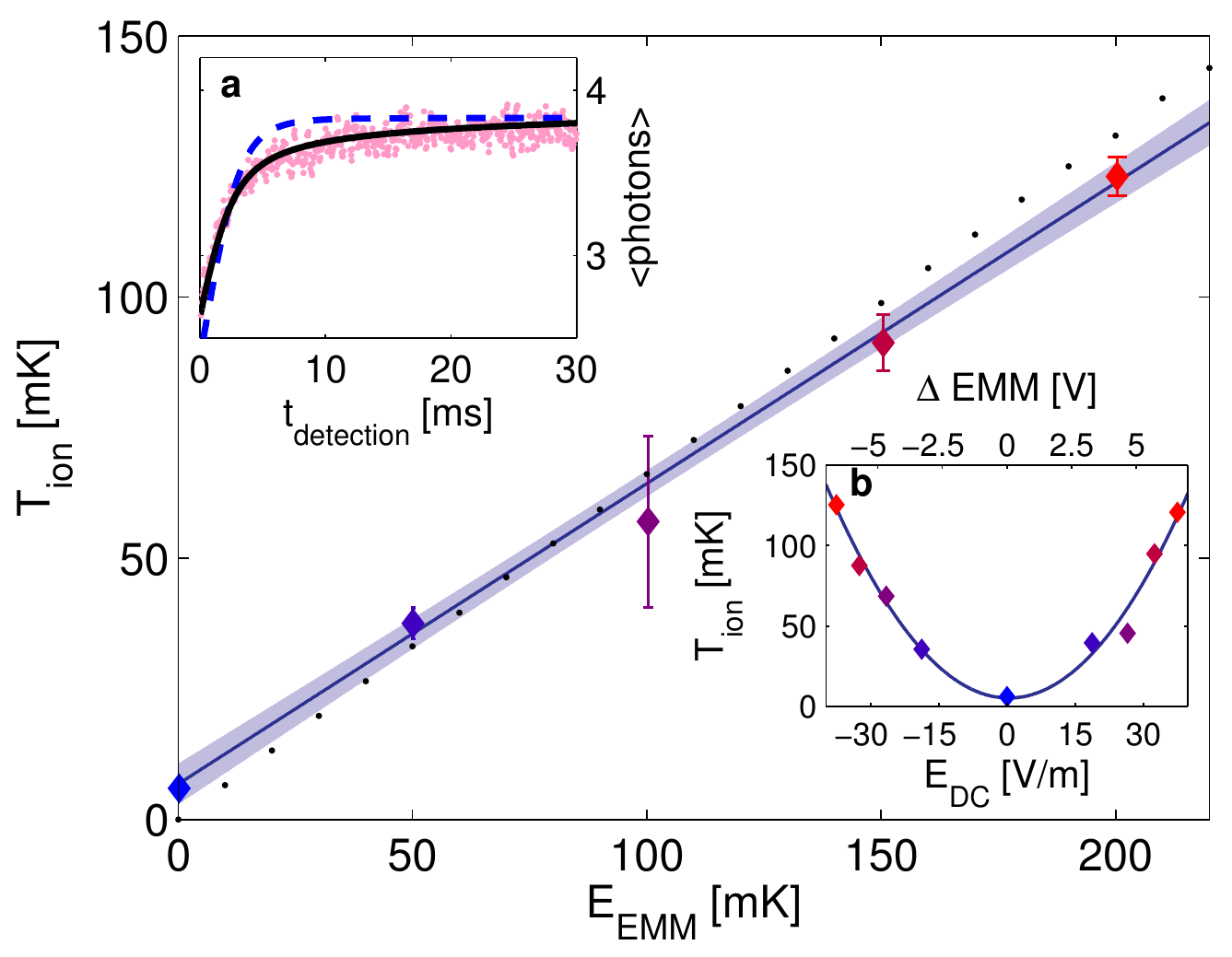}
	\caption{\textbf{Ion's steady-state temperature after 100's of atom-ion collisions.} We extract the ion's temperature from a fit to a time resolved fluorescence signal. Figure inset (a) shows a re-cooling curve (pink dots, 150$\mu$s moving average) for 150 mK$\cdot k_B$ EMM experiment and a fit for power-law (black solid line) and thermal (blue dashed line) energy distributions. Figure inset (b) shows the ion's temperature as function of the difference in the voltage on the electrode from the compensated value (top x-axis) and the resulting electric field at the ion's position (bottom x-axis). The solid blue line shown is a parabolic fit. In the main figure, the two opposing-sign DC-voltages configurations are averaged to show the ion temperature vs. the EMM kinetic energy. Error-bars accounts for both fit confidence intervals (one sigma) and the statistics of averaging over the two opposing EMM points. The solid blue line is a linear fit for the data: $T_{Ion}=0.575(19) E_{EMM}/k_B+6.8(2.4)$ mK. Shaded area represents fit confidence bounds (one sigma). The black points are the results of a simulation ($T_{Ion}=0.656 E_{EMM}/k_B$) which takes into account only the effect of hard sphere collisions.}
	\label{Fig:steadystate}
\end{figure}

The heating rates measured using Carrier Rabi spectroscopy of the narrow line-width transition show a linear increase in temperature throughout the entire measurement range (few mK). To measure the ion's temperature after longer interaction times we use DRC thermometry, which is sensitive from $\sim$10 mK to few Kelvin \cite{WesenbergRecooling}. We perform DRC using a laser slightly red-detuned (-1.8 MHz) from the $5\textrm{s}^2\textrm{S}_{1/2}\rightarrow5\textrm{p}^2\textrm{P}_{1/2}$ dipole transition and a re-pump red-detuned (-16.5 MHz) from the $4\textrm{d}^2\textrm{D}_{3/2}\rightarrow5\textrm{p}^2\textrm{P}_{1/2}$ transition. In the re-cooling analysis we take into account the eight-levels involved in the DRC, cooling of all the ion modes, radiation pressure, effects of ion micromotion and the non-thermal energy distribution of the ion \footnote{In preperation}.

To better understand the role of EMM on our ion's steady-state temperature, we scan the EMM energy by almost three orders of magnitude from 0.5 to 200 mK$\cdot k_B$. We overlap the atoms with the ion for 200 ms during which more than 400 Langevin collisions occur on average. After interaction, we detect the time-resolved fluorescence signal with 50 $\mu$s binning and up to 50 ms. As the ion cools during detection, the fluorescence signal increases. We detect on average four photons in each bin. We repeat the experiment 350 times to improve our signal-to-noise ratio. We fit the fluorescence curve to our DRC model using the power-law distribution (Eq. \ref{Eq:Tsallis}) with a single fit parameter, $T$. The power-law parameter, $n$, is fixed to the value extracted from a simulation which changes from $n$=3.9-4.2 between low and high EMM energies due to the atomic cloud finite size.
The results are shown in Fig. \ref{Fig:steadystate}.
We observe a linear dependence of the ion's temperature with the EMM energy, $T_{Ion}=0.575(19) E_{EMM}/k_B+6.8(2.4)$ mK. The scaling predicted from a simulation of hard-sphere collisions only, is $T_{Ion}=0.656 E_{EMM}/k_B$ which has a slightly higher slope, probably due to inaccuracies in DRC modeling and atomic cloud size uncertainties.
However, the main difference between the simulation and the experiment is the steady-state temperature when EMM is compensated. When EMM is compensated below 0.5 mk$\cdot k_B$, a simulation of hard-sphere collisions predicts a steady state temperature of similar magnitude (as shown in Fig. \ref{Fig:steadystate} by the crossing of the simulation data (black dots) in the origin), whereas our data indicates a steady-state temperature at least an order of magnitude higher. This is a second indication for dynamics beyond simple hard-sphere collisions.
In the figure inset we show that even the Doppler re-cooling signal is sensitive, at-least qualitatively, to the deviation of the ion's energy distribution from thermal (best fit shown by dashed line) to power-law (similarly by solid line). Here however, DRC is not sensitive enough to extract the exact power-law from the experimental data.

\begin{figure}
	\centering
	\includegraphics[width=\linewidth]{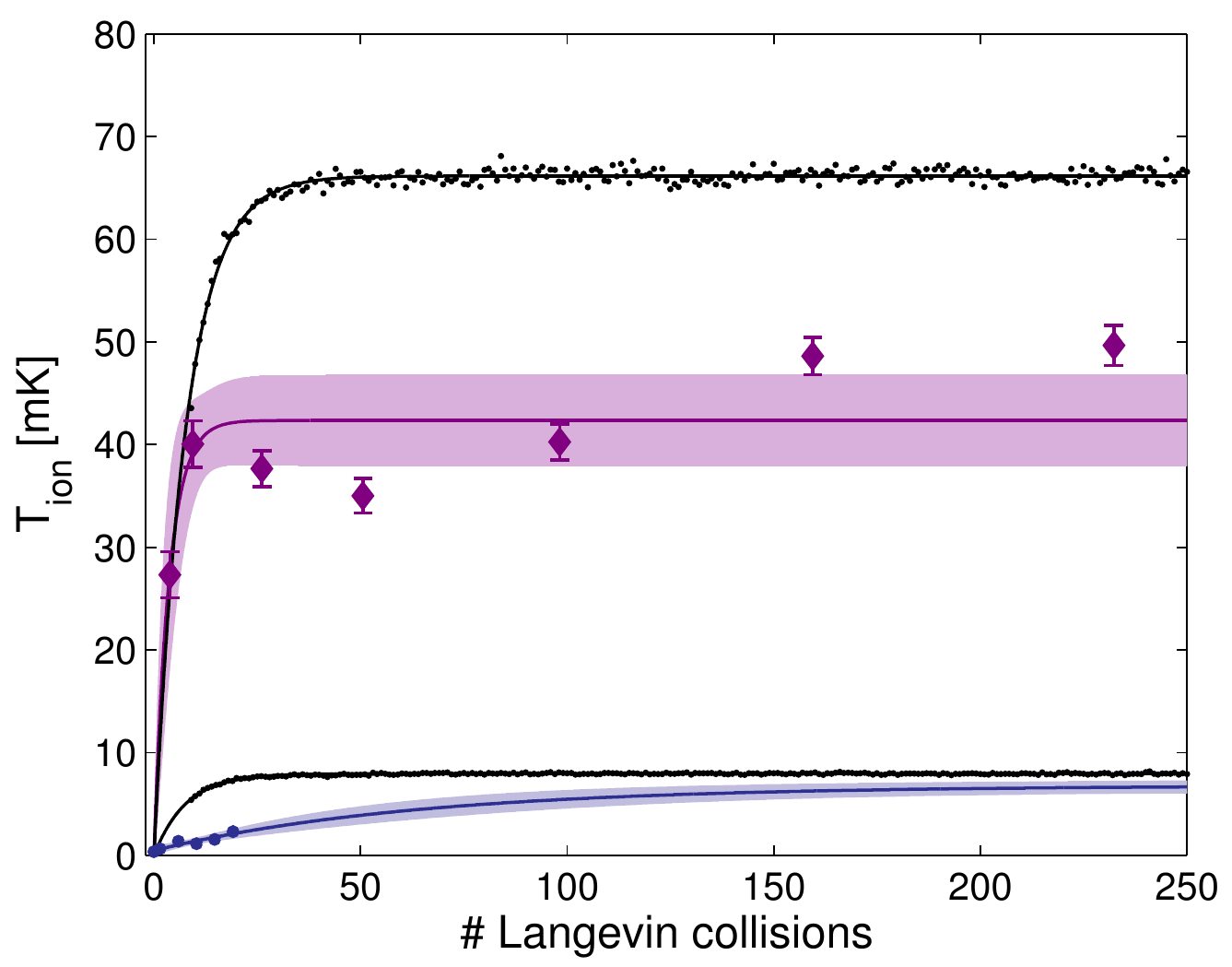}
	\caption{\textbf{Ion's heating dynamics.} Ion's temperature as it approaches steady-state for 100 mK$\cdot k_B$ EMM energy (purple diamonds) measured using DRC. From an exponential fit (purple solid line) we calculate 3.6(2.8) (1/e) collisions to reach steady-state which is with reasonable agreement with the number of collisions, 7.7, extracted from a simulation (upper set of black dots) with the same EMM energy. Error-bars are re-cooling fit confidence intervals (one sigma). Shaded area represents fit confidence bounds (one sigma). The discrepancy between the simulation and data steady-state temperature is attributed to a poor choice of EMM energy as can be seed from Fig. \ref{Fig:steadystate}b. We compare this result to the heating rate measured using Rabi spectroscopy (Fig. \ref{Fig:figure2}g, blue circles in this figure) and the steady-state measured using DRC (Fig. \ref{Fig:steadystate}, not shown in this figure) for EMM energy less than 0.5 mK$\cdot k_B$. From an exponential fit (blue solid line) we calculate 64(14) collisions (1/e) to reach steady-state. For comparison, we show a simulation results (lower set of black dots) with EMM energy (12 mK$\cdot k_B$) which results in the same steady-state temperature. Here, the number of collisions (7.7) required to reach this temperature is much smaller.}
	\label{Fig:heatingdynamics}
\end{figure}

To study the approach to steady state in the presence of EMM we measured the ion's temperature using DRC after short interaction times. The ion's temperature, as it approaches steady-state, in the presence of EMM with 100 mK$\cdot k_B$ average kinetic energy is plotted in Fig. \ref{Fig:heatingdynamics} (magenta diamonds). From an exponential fit, we extract a time-scale (1/e) of 3.6(2.8) collisions to reach steady-state. We compare this collision time-scale with a simulation for both high (\mKkB{100} as in the experiment) and low (\mKkB{12}) EMM energy (black dots) which yields a time-scale (1/e) of 7.7 collisions to reach steady-state for both. These time-scales are signature of EMM dominated collisions where the ion quickly equilibrates with the EMM.
In the absence of EMM, we observe a slow approach to steady state (64(14) collisions) which is extracted from the heating rate (100(13) $\mu$K$/$Coll.) measured using Rabi spectroscopy (Fig. \ref{Fig:figure2}g) and the steady-state (6.8(2.4) mK) measured using DRC (Fig. \ref{Fig:steadystate}).
This observation is the third indication for different dynamics in the absence of EMM.

To conclude, we used two complementary techniques to measure the ion's temperature and energy distribution after short (few collisions) and long (100s' of collisions) interaction times between a single trapped-ion, initialized in the trap ground-state and a cloud of ultra-cold atoms in the presence of negligible EMM.
Our measurements allowed us to characterize the deviation of the ion's energy distribution from Maxwell-Boltzmann to a Tsallis distribution with power-law tail. This deviation from a thermal distribution was emphasized by the use of an ion-atom mixture of nearly equal-mass species. Our system can be further used to study non-equilibrium thermodynamics.
We have seen that, in the regime of negligible EMM, ion heating is dominated by the pulling of the ion from the trap center by the atom.
Although the steady-state temperature of our ion is far from the quantum regime, the heating rate is sufficiently slow to enable us to study ultra-cold interactions by investigating the first few collisions.

This work was supported by the Crown Photonics Center, ICore-Israeli excellence center circle of light, the Israeli Science Foundation, the US-Israel Binational Science Foundation, and the European Research Council.


%

\section{Supplemental Material}

\subsection*{Apparatus}
Our apparatus consists of two vacuum chambers, connected via a thin tube. In the top chamber we collect $10^7$ $^{87}$Rb atoms in a magneto-optical-trap. The atoms are optically pumped to the F=1 hyperfine level and then loaded into a single-beam CO$_2$ (10 $\mu$m wavelength) trap where they are evaporatively cooled to temperature of $\sim$5 $\mu$K. At this stage, the atoms are loaded into a 1-D optical lattice (1064 nm YAG laser) which is used to transport the atoms to the bottom chamber. We move the atoms in the lattice by changing the relative frequency of the lattice beams by up to 3 MHz in 0.3 sec. In the bottom chamber 20,000 atoms are loaded from the optical lattice into a crossed dipole trap positioned 60 $\mu$m above the ion. We have verified that at this point the atoms are still in the F=1 hyperfine level. 
This is important due to the large hyperfine energy (325 mK) which is coupled to the ion via spin depolarization collision.

\begin{figure}
	\centering
	\includegraphics[width=\linewidth,trim={7cm 6cm 3cm 4cm},clip]{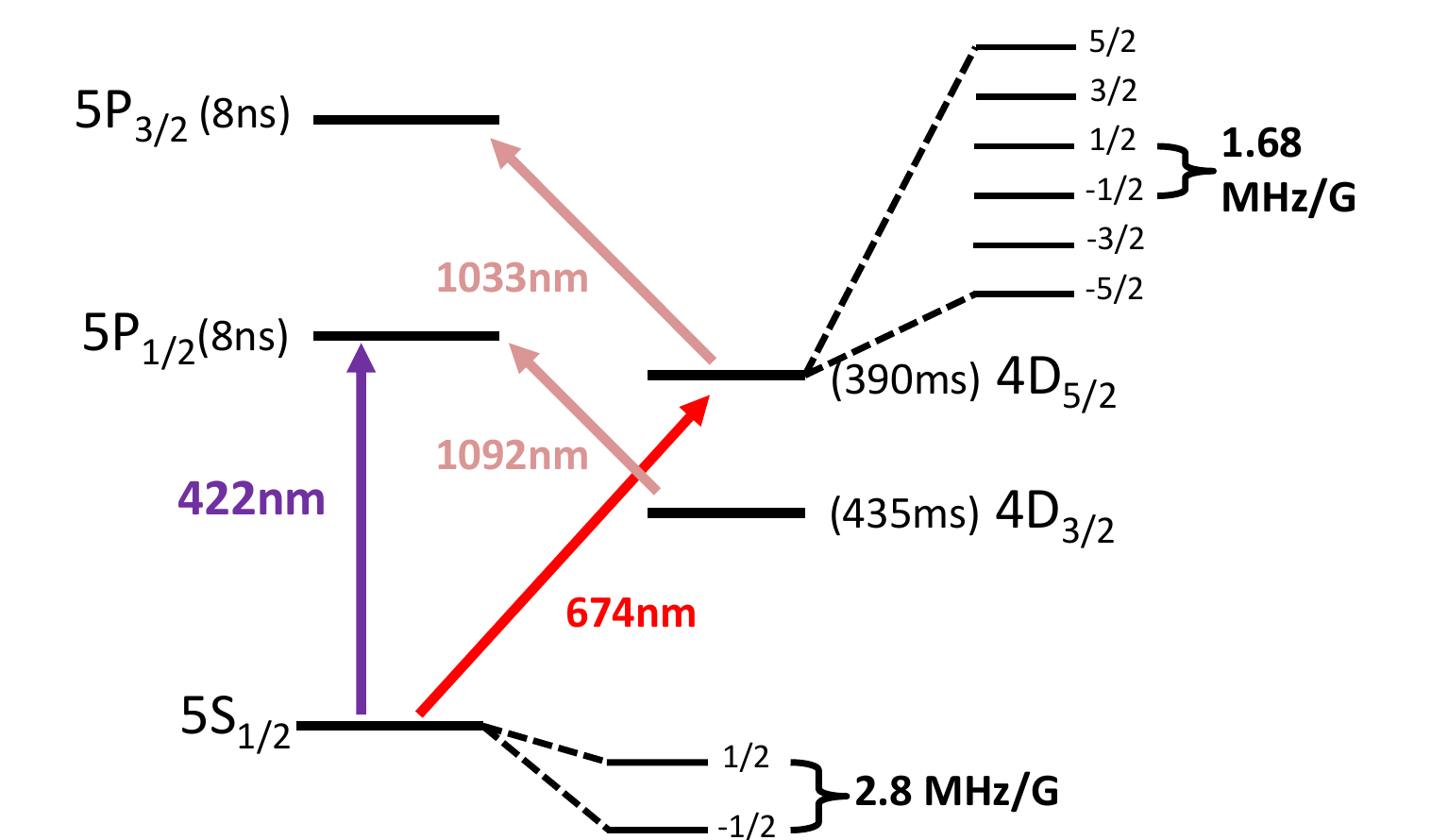}\\
	\caption{\textbf{$^{88}$Sr$^+$ ion energy levels.} We prepare the ion in the $\Sstate{-}$ electronic and Zeeman ground-state. We use a narrow line-width laser resonant with the $\textrm{S}_{1/2}\rightarrow\textrm{D}_{5/2}$ quadrupole transition at 674nm to perform coherent Rabi thermometry. We use a 422nm laser resonant with the $\textrm{S}_{1/2}\rightarrow\textrm{P}_{1/2}$ dipole transition together with a 1092nm re-pump on the $\textrm{D}_{3/2}\rightarrow\textrm{P}_{1/2}$ transition for state-selective fluorescence and Doppler re-cooling thermometry.}
	\label{Fig:figure1}
\end{figure}

At the bottom chamber a single $^{88}$Sr$^+$ ion is trapped in a linear segmented RF (26.5 MHz) Paul trap (f$_\textrm{trap}$=(0.58,0.82,1.29) MHz). We detect the ion's EMM using Rabi side-band spectroscopy on the narrow $5\textrm{s}^2\textrm{S}_{1/2}\rightarrow4\textrm{d}^2\textrm{D}_{5/2}$ quadrupole transition (see Fig. \ref{Fig:figure1} for ion's energy scheme). We use three distinct 674nm laser beams to detect EMM in all three axes. We cancel AC-magnetic field systematic shifts by using two different Zeeman transitions. We reduce the ion's total EMM kinetic energy below \mKkB{0.5} by routinely compensating EMM using both DC and RF fields on an hourly basis. This method (side-band spectroscopy) is also used to accurately determine the EMM energy in the high-EMM experiments.

\begin{figure}
	\centering
	\includegraphics[width=\linewidth]{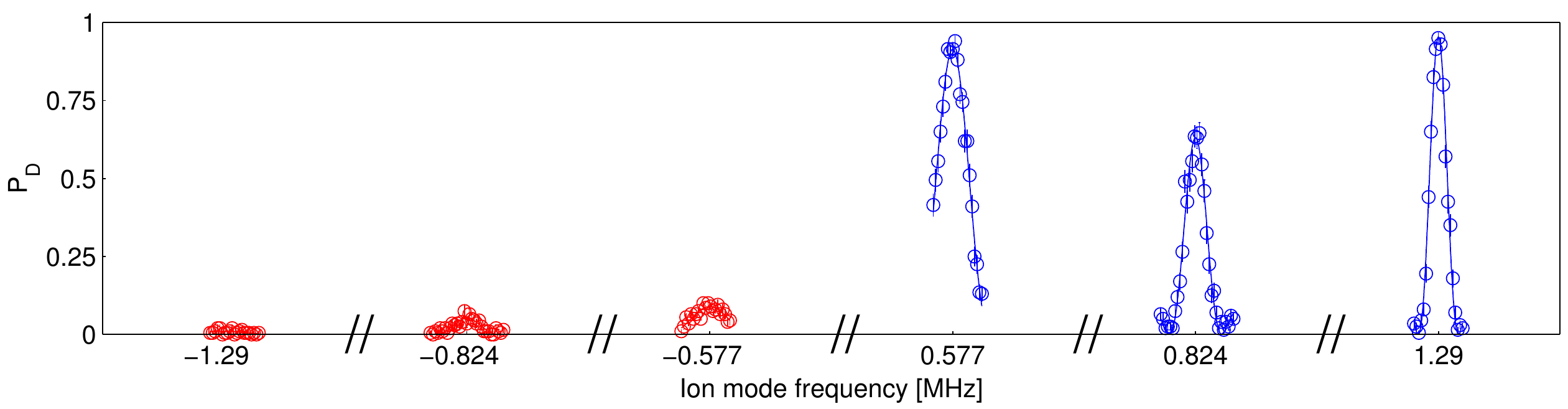}\\
	\caption{\textbf{Ion's ground-state cooling.} Rabi spectroscopy of the ion's motional side-bands (f$_\textrm{trap}$=(0.58,0.82,1.29) MHz)) showing ground-state cooling ($\bar{n}$=(0.1,0.09,0.01)) of all the ion's modes.}
	\label{Fig:GSC}
\end{figure}

While the atoms are held 60 $\mu$m above the ion, we cool the ion to its 3D motional ground-state ($\bar{n}$=0.1,0.09,0.01) and optically pump the ion to its $\Sstate{-}$ Zeeman sub level of its electronic ground state. Fig. \ref{Fig:GSC} shows a spectroscopy of the ion's motional side-bands after ground-state cooling.
To overlap the atoms with the ion, we use a piezoelectric driven mirror to move the crossed dipole trap onto the ion position in 5 ms. The atoms begin their interaction with the ion before transport is completed. To account for interaction during transport we add an equivalent extra 0.5 ms interaction time (already included in values that appear in the text). At the end of the transport the atoms oscillate with an amplitude of $\sim$3 $\mu$m for roughly 10ms. This oscillation reduce the effective density of the atomic cloud to 86$\%$ of its original value (already included in the text).
During atom-ion interaction all lasers beams are mechanically shut except for the 1064 nm used for dipole trap.
After the desired interaction time (from ms to few seconds) we release the atoms from the dipole trap and detect their number and density using absorption imaging. 
Immediately after, we perform Rabi carrier spectroscopy on the narrow $5\textrm{s}^2\textrm{S}_{1/2}\rightarrow4\textrm{d}^2\textrm{D}_{5/2}$ quadrupole transition combined with Doppler re-cooling spectroscopy on the dipole allowed $5\textrm{s}^2\textrm{S}_{1/2}\rightarrow5\textrm{p}^2\textrm{D}_{3/2}$ transition.

\subsection*{Likelihood estimation}
We perform a likelihood analysis to extract the Tsallis distribution parameters, $n$ and $T$ (Eq. 1 in the text). The likelihood function is defined as: $L\left(n,T\right)=\prod_i L_i\left(x_i;N_i,p_i\right)$ where $L_i\left(x_i;N_i,p_i\right)={{N_i}\choose{x_i}}p_i^{x_i}\left(1-p_i\right)^{N_i-x_i}$ is the likelihood to measure $x_i$ dark ion events out of $N_i$ measurements assuming that the ion's D-state population is exactly $p_i$. The index $i$ represents the different pulse times, $t_R$, in the experiment. The D-state population, $p_i\left(n,T\right)$ is determined from evaluating Eq. 2 using Tsallis distribution with given parameters, $n$ and $T$. From the likelihood function $L\left(n,T\right)$ we extract the mean and standard-deviation of the distribution parameters and also the ion's temperature (Eq. 3) and its error.

\subsection*{Simulations}
We perform two type of atom-ion collision simulations. The simpler type treats only hard-sphere collisions. Here, the motion of the ion between collisions is calculated analytically, whereas the collisions with the atoms are modeled as a stochastic process. The collision is defined by the resulting scattering angles which are randomly and uniformly sampled with each collision. Since the collisions are modeled as instantaneous, only the velocity of the ion is modified and determined by energy and momentum conservation. The time between consecutive collisions is randomly sampled from exponential distribution to render a constant rate of Langevin collisions.
In a more involved simulation we treat the ion and atoms as billiard-balls with a $-1/r^5$ attractive force. Apart from the mutual interaction, the ion also experiences the trap RF and DC fields whereas the atoms are modeled as free particles. We simulate time dynamics by letting the ion to interact with multiple atoms consecutively, i.e., one atom at a time. At t=0 the ion initial conditions are randomly sampled from a Boltzmann distribution with a temperature 0.5 mK. at the center of what we define as an interaction sphere with a radius of 1.2 $\mu$m. Next, a single Rb atom is generated in a random position on the surface of the interaction sphere with random velocity that is sampled from a Maxwell-Boltzmann distribution with temperature of 6 $\mu$K. The particles classical trajectories are calculated using the Runge-Kutta fourth-order method until the atom leaves the interaction sphere. In case of contact interaction due to the particles finite size which was set here to 5 nm, a deterministic hard sphere collision is evaluated after which the time integration continues. This process is repeated with new atoms being randomly generated whereas the ion motion is preserved between the consecutive interaction events.
In order to reduce computation time, we numerically calculate only events in which the atom-ion minimal separation is below 100 nm.
By running the simulation multiple times we obtain the energy distribution of the ion as a function of time, which we then compare to our results.
Due to the high energy tail of the energy distribution, as the ion heats, the interaction sphere of 1.2 um introduces finite size effects. Therefore, we could not use this simulation for ion temperatures well above 2 mk without an increase of the interaction sphere. Further increase resulted in too long simulation run-time.

\end{document}